\documentstyle[aps]{revtex}

\begin{document}
\begin{center} \begin{bf}
Local gauge invariance implies Siegert's hypothesis
\end{bf} \end{center}
\begin{center}
H.W.L. Naus \\
{\em Institute for Theoretical Physics, University of Hannover}\\
{\em Appelstr. 2, 30167 Hannover, Germany }
\end{center}

\begin{abstract}
The nonrelativistic Ward-Takahashi identity,
a consequence of local gauge invariance in quantum mechanics,
shows the necessity of exchange current contributions
in case of nonlocal and/or isospin-dependent potentials. 
It also implies Siegert's hypothesis: in the nonrelativistic limit,
two-body charge densities identically
vanish. Neither current conservation, which follows from 
global gauge invariance, nor 
the constraints of  (lowest order) relativity  
are sufficient to arrive at this result. Furthermore, a low-energy
theorem for exchange contributions
is established. \\
\vskip 0.1cm
PACS number(s): 23.20.-g; 25.20.-x\\
\vskip 0.1cm
\end{abstract}

\baselineskip=24.0pt
In deriving the low-energy relation between nuclear electromagnetic
current and charge multipoles, i.e. `Siegert's theorem', Siegert conjectured
that the charge density is in nonrelativistic order not modified
by exchange contributions \cite{Sieg}. This is know as Siegert's
{\it hypothesis} and it still is an essential ingredient in
explicit calculations of nuclear electromagnetic processes (see, e.g.,
ref. \cite{Aren} and references therein).
In the original paper \cite{Sieg}, using Fermi theory,
corrections have been estimated
to be of higher order in $v/c$,
with $v$ a typical velocity in the system.
As will be demonstrated below, the lowest
order relativity constraints \cite{Friar} nevertheless do not {\it a priori}
exclude a modification of the charge density. Furthermore, though
one-boson exchange does not, it is claimed \cite{Hyu} that two-boson
exchange does affect the charge density, thus conflicting Siegert's
hypothesis and violating his theorem. In this Letter it is shown that
the requirement of local gauge invariance yields Siegert's hypothesis.
In nonrelativistic quantum mechanics, with moderate restrictions on the
potentials, only the current density may be  modified.
This result, evidently in accordance with global gauge invariance,
does not follow from current conservation. 

The proof is based on the nonrelativistic Ward-Takahashi identity which
was recently derived assuming local gauge invariance \cite{Nabo}.
It relates the electromagnetic charge and current operators,
$\rho, \vec{j}$, to the
full propagator (Green's function), $G_E = (E - H)^{-1}$, of the strongly
interacting $N$-particle system:
\begin{equation}
k_0 \rho (\vec{k}) -\vec{k} \cdot \vec{j}(\vec{k}) =
G_{E'}^{-1} \rho_0(\vec{k})- \rho_0(\vec{k})G_E^{-1} \;.
\label{eq:WT}
\end{equation}
Here $k_0=E'-E$  is the energy, and $\vec{k}$ the momentum transfered
by the (virtual) photon to the system. This relation is valid for
off-shell energies $E$ and $E'$, i.e.
not corresponding to the asymptotic energies of the initial
and final states. Acting on those states, the inverse Green's functions
just yield zero.
Note that the right-hand side
of the identity contains $\rho_0$, defined as the unmodified charge density,
i.e. the sum of the one-body densities. It appears because
it generates the local gauge transformations for the particle degrees of
freedom.
In momentum space, it separately shifts
the momenta of the initial and final charged particles
by the photon momentum $\vec{k}$.

We will show now that the nonrelativistic Ward-Takahashi identity
excludes lowest order exchange contributions to the charge density, i.e.,
\begin{equation}
 \rho(\vec{k}) = \rho_0(\vec{k}) , 
\label{eq:Sie}
\end{equation}
up to order $v/c$.
Whereas the current generally has two-body pieces, the nonrelativistic
charge density is just the sum of the one-body charge densities.
Let us start with the nonrelativistic Ward-Takahashi identity 
for the noninteracting $N$-particle system  
\begin{equation}
k_0 \rho_0(\vec{k}) -\vec{k} \cdot \vec{j}_0(\vec{k}) =
(E'-T) \rho_0(\vec{k}) - \rho_0(\vec{k}) (E-T) \;,
\end{equation}
where $T$ denotes the kinetic energy operator and
$\vec{j}_0$ the unmodified current operator.
Both operators are sums of the corresponding one-body operators.
We include interactions via $H=H_0+V=T+U+V$;  $U$
is defined as that part of the potential which commutes
with the charge density, i.e. $[U, \rho_0] =0 $. 
As in ref. \cite{Nabo}, time
(energy) dependent interactions are excluded. One readily
verifies that the unmodified charge and
current operators also fulfil 
\begin{equation}
k_0 \rho_0(\vec{k}) -\vec{k} \cdot \vec{j}_0(\vec{k}) =
(E'-H_0) \rho_0(\vec{k}) - \rho_0(\vec{k}) (E-H_0) \;.
\label{eq:WT0}
\end{equation}
In case there are no other interactions ($V=0$),
local gauge invariance does not require two-body currents and
eq. (\ref{eq:WT}) is satisfied. If  interactions
are present which do  not commute with $\rho_0$, for instance
because of nonlocalities and/or isospin dependencies in $V$, exchange
contributions are necessary. This immediately can be seen
from the nonrelativistic Ward-Takahashi identity, rewritten by means
of eq. (\ref{eq:WT0}),  
\begin{equation}
(E'-E) \rho_e(\vec{k}) -\vec{k} \cdot \vec{j}_e(\vec{k}) =
[ \rho_0(\vec{k}), V] \;,
\label{eq:WTe}
\end{equation}
where $\rho_e=\rho -\rho_0 \, , \; \vec{j}_e=\vec{j}-\vec{j}_0$.
Furthermore, since none of the appearing operators depends on energy and
the relation is valid for arbitrary (off-shell) energies
$E$ and $E'$, it immediately implies that $\rho_e = 0$
up to order $v/c$.
Thus Siegert's hypothesis, cf. eq. (\ref{eq:Sie}),
has been proven from the nonrelativistic
Ward-Takahashi identity, eq. (\ref{eq:WT}).

In this nonrelativistic framework with time-independent potentials
the energy dependence in eq. (\ref{eq:WT}) is rather trivial and,
as a consequence, $E (E')$ only appears explicitly in eq. (\ref{eq:WTe}).
This breaks down in a covariant approach, where
analogous `generalized Ward identities' have been derived 
by Kazes \cite{Kazes} a long time ago. However, due to the energy-dependence
of the operators involved, a similar conclusion about
the two-body charge density cannot be drawn.
Nevertheless, these relations can be exploited in, for example,
deriving low-energy theorems (see, e.g., ref. \cite{nfk}).

The off-energy shell implementation of
electromagnetic current conservation, 
\begin{equation}
\vec{k} \cdot \vec{j} (\vec{k}) = [ H, \rho(\vec{k})]  \;,
\label{eq:CC}
\end{equation}
which recently appeared in the literature 
\cite{FF}, reads
\begin{equation}
k_0 \rho (\vec{k}) -\vec{k} \cdot \vec{j}(\vec{k}) =
G_{E'}^{-1} \rho(\vec{k}) - \rho(\vec{k}) G_E^{-1} \;.
\label{eq:WTFF}
\end{equation}
The on-shell matrix element of the right-hand side of this relation
also reduces to zero.
Strictly speaking, local gauge invariance is not required here;
a {\it global}
symmetry already guarantees a conserved current. Since the electromagnetic
field couples to a conserved current, eq. (\ref{eq:WTFF}) should
hold, irrespective of requiring a local $U(1)$ symmetry. In other words,
eq. (\ref{eq:WT})  is a stronger constraint than eq. (\ref{eq:WTFF}).
Simultaneously demanding both relations, i.e. the nonrelativistic
Ward-Takahashi identity
and current conservation, indeed indicates that the charge density is not
modified by exchange contributions.

It is also instructive to consider the equation for the two-body
current only imposing current conservation. Instead of eq. (\ref{eq:WTe})
one readily  obtains
\begin{equation}
[ H, \rho_e(\vec{k})] -\vec{k} \cdot \vec{j}_e(\vec{k}) =
[ \rho_0(\vec{k}), V] \;.
\label{eq:FFe}
\end{equation}
Eqs. (\ref{eq:WTe}) and (\ref{eq:FFe}) 
reduce to the same matrix element equation with respect to
eigenstates of the Hamiltonian $H$. 
Obviously, as operator equations they are not equivalent.
In particular, Siegert's hypothesis {\it only} follows from eq. (\ref{eq:WTe}),
i.e. local gauge invariance.
Indeed schemes for extracting exchange current contributions
compatible with current conservation, cf. eqs.
(\ref{eq:CC}, \ref{eq:WTFF}, \ref{eq:FFe}),
however modifying the charge density have been developed  \cite{Hyu}.
Consequently, these currents violate the nonrelativistic
Ward-Takahashi identity.
Other extraction schemes, see e.g. \cite{Henn}, respect
this identity, i.e. two-body charge densities vanish
in the nonrelativistic limit.

It is important to realize that, because of the use of
the electromagnetic four-potential $A_{\mu}$,  one cannot defer
relativity completely even in  noncovariant approaches like in ref. \cite{Nabo}.
There it is of course tacitly assumed that the electromagnetic coupling
is $A_{\mu} j^{\mu}$ and therefore
$j^{\mu} (\vec{x})=(\rho (\vec{x}), \vec{j} (\vec{x}))$ 
should transform as a four-vector
\cite{Friar,Close}. In terms of the `boost' operator $\vec{K}$
one has the conditions
\begin{eqnarray}
[ \vec{K}, \rho (0) ] & = & i \; \vec{j}(0) \, , \nonumber\\
 \left[ K^{m} , j^{n} (0) \right] & = & i \; \delta^{nm} \rho (0).
\label{eq:boost}
\end{eqnarray}
The operator $\vec{K}$ can be expanded in $v/c$ (or $1/m$,
where $m$ is the mass of the particles)
\cite{Friar,Close}. For our purposes we only need the lowest order term,
\begin{equation}
\vec{K} \simeq \vec{K}_0 = M \vec{R} \; ,
\end{equation}
with the CM-coordinate $\vec{R}$ and the total mass $M$.
Based on eq. (\ref{eq:boost}), Friar gave a classification
of exchange vector and axial currents \cite{Friar}.  Recall
the usual realization in case of the vector current: $\rho_e$
is of the same order as relativistic corrections to $\rho_0$, i.e. $(v/c)^2$.
On the other hand, the exchange contribution $\vec{j}_e$ is of
the same order as $\vec{j}_0$; both are $O(v/c)$, whereas relativistic
corrections are of order $(v/c)^3$. This is actually  compatible
with the constraints above, cf. eq. (\ref{eq:boost}), provided that
\begin{equation}
 \left[ K^{m}_{0} , j^{n}_{e} (0) \right]  = 0 \; .
\end{equation}
However, the conditions (\ref{eq:boost})
also allow for a {\it different}
realization: if 
\begin{equation}
[ \vec{K}_0, \rho_e (0) ]  =  0 \, , 
\end{equation}
then the exchange charge density,
$\rho_e$, can be of nonrelativistic order, i.e. $O(1)$.
In other words, based on the lowest order relativity constraints,
one cannot {\it a priori} exclude two-body charge density contributions.
Local $U(1)$ symmetry of the Schr\"{o}dinger theory does exclude these.

An immediate consequence of this work is that it proves
the low-energy theorem proposed in ref. \cite{Nabo}.
Consider the two-particle system, $e_1=e, e_2=0$, interacting via a nonlocal
potential $V$. In lowest order of the photon momentum the exchange
current is unambiguously given by
\begin{equation}
\vec{j}_e(\vec{k}) = e \frac{m_1}{m_1+m_2} \left[
\nabla_{\vec{p}} \, {\cal V}(\vec{p}\, ', \vec{p})
+ \nabla_{\vec{p} \,'}\,  {\cal V}(\vec{p} \,', \vec{p}) \right],
\end{equation}
where $\cal V$ denotes the matrixelement of $V$ in terms of relative
momenta after separating the CM-motion;
$m_1 $ and $m_2$ are the respective masses.
Since the two-body charge density necessarily vanishes, this yields an
unique prediction for $( \rho, \vec{j})$ in lowest order of $\vec{k}$.
Note that the lowest order two-body current is purely
longitudinal. Consequently, it does not contribute for real photons.

In this Letter we proved that Siegert's hypothesis follows from
the requirement of local gauge invariance in nonrelativistic quantum
mechanics. We believe that the gauge principle is rather universal,
i.e. applicable to  `fundamental' as well as
`effective' theories and, appropriately implemented, in classical
as well as quantum mechanical formulations.
Maxwell's equations, for instance,
seem to apply well in the microscopic domain where
the charge and current densities are treated quantum mechanically
\cite{Jack}. Given this observation, one indeed then can introduce
the electromagnetic potentials, necessarily leading to
local gauge invariance. 
In the quantum mechanical realization of this gauge symmetry,
via the generator of the local phase transformations, the 
nonrelativistic Ward-Takahashi identity is valid.
Nevertheless, an alternative view is possible.
Local gauge invariance is supposed to be imposed in some underlying
(more) fundamental theory, from which an effective model should be derived.
The actual construction, however,
usually not only involves approximations but
also may be rather decoupled from the
underlying theory. For instance, in introducing realistic nucleon-nucleon
potentials  a clear connection to QCD is lacking. 
One therefore may argue that imposing current conservation  is
sufficient and that local gauge invariance is too strong a condition.
Clearly, this does not cover the point of view that constraints
based on symmetries like $U(1)$ local invariance are useful 
in constructing effective models. These constraints, e.g.
Ward-Takahashi identities, also provide consistency checks
for practical calculations in such models. Our findings in this work,
in particular the theoretical confirmation of the phenomenologically
well-established Siegert hypothesis,
support this latter point of view.

\bigskip The author would like to thank J.L. Friar, J.H. Koch and P.U. Sauer for
useful discussions and/or a critical reading of the manuscript.

\end{document}